\documentclass[twocolumn,showpacs,preprintnumbers, 
               superscriptaddress]{revtex4}
\usepackage{graphicx, fancybox}
\usepackage{amsmath,amssymb}

\newcommand{\bm}[1]{\mbox{\boldmath$#1$}}

\begin{document}

\preprint{RBRC-620}

\title{
Chiral transition and mesonic excitations 
for quarks with thermal masses
}

\author{Yoshimasa Hidaka}
\email{hidaka@quark.phy.bnl.gov}
\affiliation{
RIKEN-BNL Research Center, Brookhaven National Laboratory, Bldg.510A, Upton, 11973, NY, USA}

\author{Masakiyo Kitazawa}
\email{kitazawa@quark.phy.bnl.gov}
\affiliation{
RIKEN-BNL Research Center, Brookhaven National Laboratory, Bldg.510A, Upton, 11973, NY, USA}

\begin{abstract}

We study the effect of a thermal quark mass, $m_T$, 
on the chiral phase transition and mesonic excitations
in the light quark sector at finite temperature 
in a simple chirally-symmetric model.
We show that while nonzero $m_T$ lowers the chiral condensate,
the chiral transition remains of second order.
It is argued that the mesonic excitations have large decay rate
at energies {\it below} $2m_T$, owing to the Landau damping of 
the quarks and the van Hove singularities of the collective modes.

\end{abstract}

\date{October 27, 2006}

\maketitle

\section{Introduction}

Quantum chromodynamics (QCD) at nonzero temperature
exhibits rich physics, and
is expected to undergo chiral and deconfinement 
transitions at a critical temperature $T_\text{c}$.
The RHIC experiments revealed interesting features of
a quark-gluon plasma (QGP) phase above $T_\text{c}$ 
\cite{RHIC}.
One of the historical problems is
to explore the existence and the nature of the mesonic excitations 
in the QGP phase \cite{HK85,J/Psi,Lattice,Shuryak:2003ty}.
Recent lattice QCD simulations suggest the existence of
such excitations
in the heavy quark sector \cite{Lattice}.
In the light quark sector, it is theoretically predicted
that there exist light collective modes
in the sigma- and pi-channels, as the soft modes of
the chiral phase transition \cite{HK85}.

It is also known that the quarks show a peculiar behavior
in the QGP phase.
At extremely high temperatures,
the hard thermal-loop (HTL) approximation \cite{HTL}
can be adopted, and
the quark propagator has two collective excitations, 
normal quasiparticles and plasmino 
modes \cite{HTL,plasmino,LeBellac}.
The excitation spectra of both collective modes have 
a mass gap termed the ``thermal mass''
$m_T \sim gT$, where $g$ and $T$ are the gauge coupling
and the temperature, respectively.
Unlike the Dirac mass, the thermal mass does not
break the chiral symmetry.
Although the validity of the HTL approximation may be violated 
around $T_\text{c}$, 
lattice simulations suggest a large quark mass 
near but above $T_\text{c}$ \cite{Petreczky:2001yp}.
If the quarks have large thermal masses near $T_\text{c}$,
they can affect the properties of mesonic excitations, 
especially in the light quark sector \cite{Shuryak:2003ty}, 
and also the chiral transition at finite temperature.
We notice that the quark spectrum can have a thermal mass
in the vicinity of $T_\text{c}$ as a result of the soft mode of the
chiral transition \cite{Kitazawa:2005mp}.

In the present work,
we explore the effects of the thermal mass 
on the chiral phase transition and mesonic excitations 
at finite temperature in a simple model.
We employ the propagator 
obtained in the HTL approximation for the quarks, 
with the thermal mass $m_T$ introduced as a parameter. 
We also adopt a chirally-symmetric four-Fermi interaction 
in the scalar and pseudoscalar channels.
Using this model, we show that the order of the chiral transition
does not change, while the value of the chiral condensate 
is suppressed, when $m_T \ne 0$ is included.
It will be shown that even at finite $m_T$, 
there are soft modes of the chiral
transition in the scalar and pseudoscalar
channels \cite{HK85}.
Using these modes, we study the effect of $m_T$ on
the mesonic excitations above $T_\text{c}$.

How does the thermal mass of the quark affect 
the properties of mesonic excitations?
Naively, 
a mesonic excitation below the expected threshold energy,
 $\omega_{\rm thr}=2m_T$, 
is tightly bounded,
since the decay process into quark and anti-quark is 
forbidden below $\omega_{\rm thr}$.
We show in the present work, however, that
the effect of the thermal mass is completely contrary to
this statement;
if the quark has the spectrum calculated in the HTL approximation
with a thermal mass $m_T$,
the decay rate of the mesonic excitations is enhanced
below the energy $2m_T$.
It is discussed that these decay rates mainly originate from
the continuum in the quark spectrum which physically corresponds
to the Landau damping of the quasi-quarks.

This paper is organized as follows.
In the next Section, we present the model employed in this work.
In Section III, we calculate the chiral condensate 
in the mean-field approximation 
and demonstrate the chiral phase transition in our model.
We then calculate the spectral function in the scalar and
pseudoscalar channels above $T_\text{c}$ in Section IV.
Section V is devoted to discussions and a summary.

\section{Model} \label{sec:formalism}

Before we present the model employed, 
let us first briefly review the property of the quark propagator
in the high temperature limit of QCD, in order to see
the emergence of the gapped collective excitations 
in the quark spectra.
In this limit, 
the retarded quark propagator is calculated by
the HTL approximation as
\begin{align}
S^{\rm HTL}( \omega,\bm{p} )
&= \left[ (\omega + i\eta)\gamma^0 - \bm{p}\cdot\bm{\gamma} 
- \varSigma^{\rm HTL} ( \omega+i\eta, \bm{p} ) \right]^{-1},
\label{eq:S}
\end{align}
where $p=|\bm{p}|$ and
\begin{align}
\varSigma^{\rm HTL} ( \omega,\bm{p} )
= \frac {m_T^2}p Q_0 \left( \frac\omega{p} \right) \gamma^0 
+ \frac {m_T^2}p  (1 - \frac{\omega}{p} Q_0 \left( \frac\omega{p} \right)
) \bm{\gamma}\cdot \hat{\bm{p}},
\label{eq:varSigma}
\end{align}
is the quark self-energy in the one-loop level
with $m_T ^2  = (1/6) g^2 T^2$ and
$ Q_0 = (1/2) \ln (x+1)/(x-1)$ \cite{LeBellac}.
Notice that the quark propagator $S^{\rm HTL}( \omega,\bm{p} )$
is chirally symmetric
as one can easily check that it anti-commutes with $\gamma^5$.
The quark propagator Eq.~(\ref{eq:S}) can be decomposed as 
\begin{align}
S^{\rm HTL} ( \omega,\bm{p} )
= ( S^{\rm HTL}_+ ( \omega,\bm{p} ) P_+(\bm{p}) 
+ S^{\rm HTL}_- ( \omega,\bm{p} ) P_-(\bm{p}) ) \gamma^0,
\label{eq:S^HTL_pm}
\end{align}
with projection operators 
$P_\pm(\bm{p}) = ( 1  \pm \gamma _0 \bm{\gamma }\cdot \bm{\hat p} )/2$.
The spectral functions corresponding to 
$S^{\rm HTL}_\pm( \omega,\bm{p} ) $ are found to be,
\begin{align}
\rho _ \pm^\text{HTL}  (\omega ,\bm{p}) 
\equiv &-2\text{Im} S^\text{HTL}_{\pm}(\omega,\bm{p})\nonumber\\
=& 2\pi [Z_\pm(p)\delta (\omega  - \omega _ \pm  (p)) 
       + Z_\mp(p)\delta (\omega  + \omega _ \mp  (p))]\nonumber\\
 &+ \rho^\text{L} _ \pm  (\omega ,\bm{p}).
\label{eq:quarkspectral}
\end{align}
In this limit, $\rho^\text{HTL}_+( \omega,\bm{p} )$ and 
$\rho^\text{HTL}_-( \omega,\bm{p} )$ have two poles at
$\omega=\pm \omega_\pm(p)$ and $\pm\omega_\mp(p)$, respectively,
with $\omega_\pm(p)>0 $ and 
the residues $Z_\pm(p) =({{\omega_{\pm}^2(p)  - p^2 }})/(2m_T^2)$
 \cite{LeBellac}.
Since the excitation spectra $\omega_\pm(p)$ becomes $m_T$ 
at $p=0$, $m_T$ is called the thermal mass.
The spectral functions Eq.~(\ref{eq:quarkspectral}) have 
the continuum $\rho^\text{L}_\pm( \omega,\bm{p} )$ in the space-like region,
which
originates from the Landau damping of the quark \cite{LeBellac}.
Since the quark propagator Eq.~(\ref{eq:S}) is a function of
$\omega$, $\bm{p}$ and $m_T$, in the following we write
$S^{\rm HTL}_{m_T} ( \omega,\bm{p} ) \equiv 
S^{\rm HTL} ( \omega,\bm{p} )$
to show $m_T$ dependence explicitly.

While the HTL and one-loop approximations used in the above discussion
may be invalid near $T_\text{c}$,
numerical results from lattice QCD obtained
large masses near $T_\text{c}$ \cite{Petreczky:2001yp}.
It is therefore plausible that the quarks have 
large thermal masses 
in the non-perturbative region near $T_\text{c}$.
The non-perturbative gauge interactions 
also produce an attractive interaction between quarks and 
antiquarks, and generate the chiral symmetry breaking.
In order to model this possibility,
we adopt Eq.~(\ref{eq:S}) for the quark propagator near $T_\text{c}$ and
introduce a four-Fermi quark-antiquark interaction.
We then arrive at the following model;
\begin{align}
\mathcal{L} 
= \bar \psi [S^\text{HTL}_{m_T}]^{-1} \psi  
+ G_\text{S}[(\bar \psi \psi )^2  + (\bar \psi i\gamma _5 \psi )^2 ],
\label{eq:L}
\end{align}
where $\psi$ is the quark field 
in the chiral limit,
and $G_\text{S}$ is the scalar coupling.
The thermal mass $m_T$ is introduced as a parameter 
to be varied by hand.
We introduce the three-dimensional cutoff $\varLambda$ to eliminate
the ultraviolet divergence.
This model with $m_T=0$ is equivalent to
the Nambu--Jona-Lasinio model \cite{NJL}.

\section{Chiral phase transition} \label{sec:chiral}

Let us first explore the chiral phase transition 
at finite temperature in the model of Eq.~(\ref{eq:L}).
In the mean-field approximation,
the chiral condensate $ \sigma \equiv -2G_\text{S} \langle \bar\psi\psi \rangle $
is determined by minimizing the free energy 
\begin{align}
V(\sigma) 
=& 
\frac{ \sigma^2 }{ 4G_\text{S} } - T\sum\limits_n 
\int \frac{ d^3 k} {(2\pi )^3 }
{\text{tr}}\ln [\tilde{S}( i\omega_n , \bm{k}; m_T,\sigma )]^{-1}
\nonumber \\
=&
\frac{ \sigma^2 }{ 4G_\text{S} } 
- \int \frac{ d^3 k} {(2\pi )^3 }
\int \frac{ d\omega }{\pi}\tanh\frac{\omega}{2T}\nonumber\\
&\hspace{1.2cm}\times{\text{tr}} {\rm Im} \ln [S^\text{R}( \omega , \bm{k}; m_T,\sigma )]^{-1},
\label{eq:effpot}
\end{align}
where 
$\omega_n=(2n+1)\pi T$ is the Matsubara frequency for fermions and
\begin{eqnarray}
\tilde{S}( i\omega_n,\bm{k};m_T,\sigma ) 
=  \left( [ \tilde{S}^{\rm HTL}_{m_T}( i\omega_n , \bm{k}) ]^{-1}
 - \sigma \right)^{-1},
\end{eqnarray}
with $\tilde{S}^{\rm HTL}_{m_T}$ being the Matsubara propagator
of the quark in the HTL approximation;
$\tilde{S}^{\rm HTL}_{m_T} ( i\omega_n,\bm{k} ) 
= S^{\rm HTL}_{m_T} ( \omega,\bm{k} )|_{\omega\to i\omega_n}$.
The retarded propagator $S^\text{R}$ in the right hand side 
of Eq.~(\ref{eq:effpot}) is defined by the analytic continuation
$S^\text{R}(\omega) = \tilde{S}(i\omega_n)|_{i\omega_n \to \omega+i\eta}$.
The global minimum of Eq.~(\ref{eq:effpot}) satisfies
the following stationary condition (gap equation);
\begin{align}
\frac {-1}{2G_\text{S}} 
=& \int \frac{ d^3 k} {(2\pi )^3 }
\int \frac{ d\omega }{\pi}\tanh\frac{\omega}{2T}
{\rm tr} {\rm Im}S^\text{R} ( \omega, \bm{k} ; m_T, \sigma ).
\label{eq:gapeq}
\end{align}
In general, Eq.~(\ref{eq:gapeq}) has several solutions
corresponding to local extrema of Eq.~(\ref{eq:effpot}).
One must, thus, refer to Eq.~(\ref{eq:effpot})
to find the global minimum.
In our case, however, we have checked that 
Eq.~(\ref{eq:gapeq}) always has only one solution and it is 
certainly the minimum of Eq.~(\ref{eq:effpot}).
Therefore, we determine the order parameter 
only with Eq.~(\ref{eq:gapeq}).

\begin{figure}[tbp]
\begin{center}
\includegraphics[width=.49\textwidth]{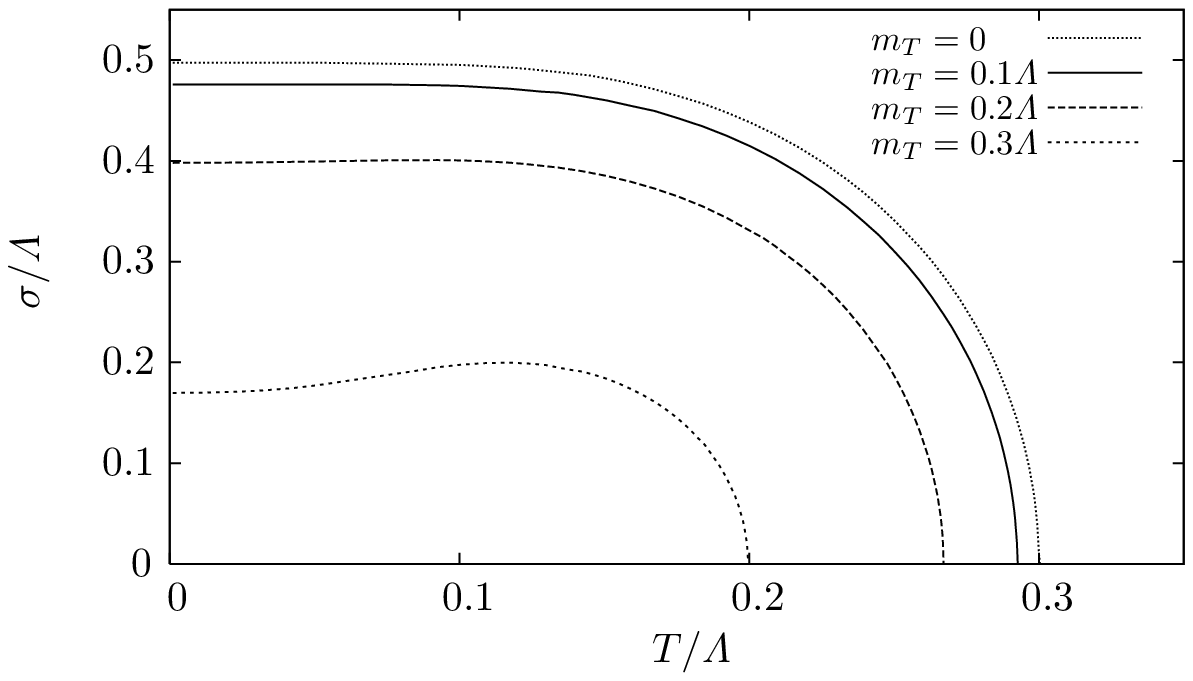}
\caption{
Order parameters for the chiral phase transition
$ \sigma \equiv -2G_\text{S} \langle \bar\psi\psi \rangle $
for $m_T/\varLambda = 0 , 0.1 , 0.2$ and $0.3$ 
with fixed $G_\text{S} \varLambda^2 = 13.01$.
The chiral transition is of second order irrespective of $m_T$.
For larger $m_T$, $\sigma$ becomes smaller.
}
\label{fig:orderp}
\end{center}
\end{figure}

In Fig.~\ref{fig:orderp}, we show the chiral condensate $\sigma$ 
as a function of $T$
with several values of $m_T$;
although $m_T$ may change with a variation of $T$, 
we fix it to study the effect of $m_T$.
The scalar coupling $G_\text{S} \varLambda^2 = 13.01$ is fixed 
so that $T_\text{c}/\varLambda = 0.3$ at $m_T=0$.
From the Figure, we see that the chiral phase transition is
of second order for any $m_T$. 
This result is natural in the sense of the symmetry argument 
of the phase transitions, 
since the symmetry of the system does not change with the 
incorporation of $m_T$.
It is notable, however, that
the order parameter $\sigma$ no longer has the meaning of
the mass gap of the quark spectra with finite $m_T$.

One sees that, as $m_T$ becomes larger, 
the chiral condensate decreases and similarly 
the critical temperature does.
This behavior is physically interpreted as the effect of
the excitation energy $m_T$ in the quark spectra:
Because the chiral symmetry breaking is induced by the condensate 
between the quark and anti-quark, the increase of the excitation energy
of these particles suppresses the condensate.
Here, we remark that
if we incorporate the current quark mass $m_\text{c}$ 
in our model, 
the value of the chiral condensate becomes larger \cite{NJL},
although $m_\text{c}$ also causes a mass gap in the quark spectra.
This is because $m_\text{c}$ 
explicitly breaks the chiral symmetry and works as an external field;
in terms of the effective potential $V(\sigma)$, 
$m_\text{c}$ gives rise to a linear term proportional to $m_\text{c} \sigma$.

\section{Mesonic excitations} \label{sec:softmode}

Since the chiral transition is of second order in our model,
there should appear soft modes associated with the transition 
\cite{HK85}.
In this Section, 
we show that this is indeed the case.
Using these soft modes, we then discuss the effect of the 
thermal mass on the mesonic excitations above $T_\text{c}$.
In the following, we limit our attention 
in the chirally symmetric phase above $T_\text{c}$,
where the excitation spectra in the scalar and pseudoscalar
channels are degenerate.

\begin{figure}[t]
\begin{center}
\begin{align}
D^\text{R}_\sigma =& G_\text{S} + 
\parbox{12mm}{\includegraphics[width=12mm]{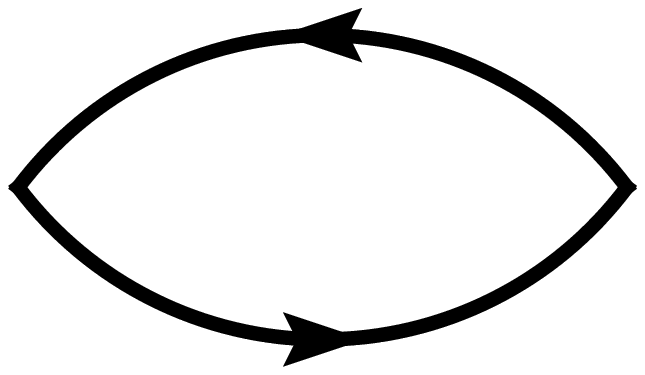}} + 
\parbox{24mm}{\includegraphics[width=12mm]{polarization2.eps}\hspace{-0.1cm}
\includegraphics[width=12mm]{polarization2.eps}} + \cdots 
\nonumber \\
=& \frac{-1}{ G_\text{S}^{-1} + \varPi^\text{R}_\sigma }
\nonumber \\
\varPi^\text{R}_\sigma = &
\parbox{12mm}{\includegraphics[width=12mm]{polarization2.eps}} 
=
\parbox{15mm}{\includegraphics[width=15mm]{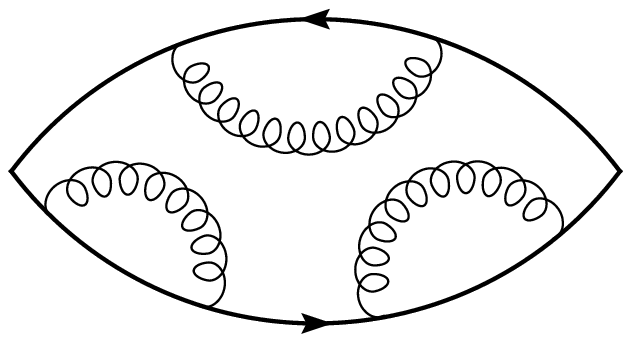}} + \cdots,
\hspace{4mm}
\parbox{12mm}{\includegraphics[width=12mm]{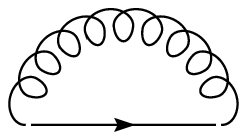}} 
= \varSigma^{\rm HTL}_{m_T}
\nonumber
\end{align}
\caption{
Diagrammatic representation for the meson propagator 
$D^\text{R}_\sigma(\omega,\bm{p})$ and the polarization function 
$\varPi^\text{R}_\sigma(\omega,\bm{p})$.
The thick-solid line denotes the quark propagator
in the HTL approximation Eq.~(\ref{eq:S}), while
the thin-solid line represents the free quark propagator.
}
\label{fig:resum}
\end{center}
\end{figure}

In the random phase approximation, 
the retarded meson propagators in these channels 
$D_\sigma^\text{R}(\omega,\bm{p})$ is given by
\begin{eqnarray}
D_\sigma^\text{R} ( \omega ,\bm{p})
= \frac{-1}{ G_\text{S}^{-1} + \varPi_\sigma^\text{R}( \omega , \bm{p} ) },
\label{eq:D^R}
\end{eqnarray}
where the polarization function $\varPi_\sigma^\text{R}(\omega,\bm{p})$ 
is determined by the analytic continuation 
$\varPi_\sigma^\text{R}(\omega,\bm{p}) 
= \tilde\varPi_\sigma( i\nu_n,\bm{p} )|_{i\nu_n\to\omega+i\eta}$
with 
\begin{align}
\tilde\varPi_\sigma( i\nu_n,\bm{p} )
=&
2T\sum_m\int\frac{ d^3k}{(2\pi)^3}
{\text{tr}} [
S^\text{HTL}_{m_T}(i\omega_m,\bm{k}) \nonumber\\
& \times
S^\text{HTL}_{m_T}(i\nu_n+i\omega_m,\bm{p}+\bm{k}) ]
\nonumber \\
=&
2 \sum_{s,t=\pm} \int\frac{d^3k}{(2\pi)^3}
{\rm tr} [ P_s(\bm{k}) \gamma^0 P_t(\bm{p}+\bm{k}) \gamma^0 ]\nonumber\\
&\times\int \frac{ d\omega_1 d\omega_2 }{ (2\pi)^2 }
\frac{ f(\omega_1) - f(\omega_2) }{ i\nu_n + \omega_1 - \omega_2 }\nonumber\\
&\times\rho^{\rm HTL}_s ( \omega_1,\bm{k} )
\rho^{\rm HTL}_t ( \omega_2,\bm{p}+\bm{k} )
\label{eq:Pi}
\end{align}
and the Matsubara frequency for bosons $\nu_n = 2n\pi T $.
In Fig.~\ref{fig:resum}, we show the diagrammatical representation
of our approximation.

\begin{figure} 
\begin{tabular}{cccc}
\parbox{.5cm}{(a)\vspace{.9cm}} &
\parbox{2cm}{\includegraphics[width=.11\textwidth]{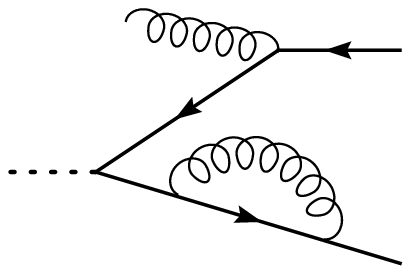}} & \hspace{5mm}
\parbox{.5cm}{(b)\vspace{.9cm}} &
\parbox{2cm}{\includegraphics[width=.11\textwidth]{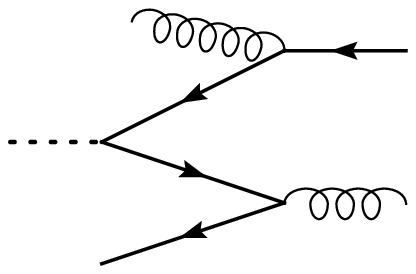}}
\end{tabular}
\caption{
The diagrammatical interpretation for the decay processes 
included in $I_\text{PC}(\omega)$ (left) and $I_\text{CC}(\omega)$ (right).
The dashed line denotes the mesonic excitations and
the temporal direction is left to right.
}
\label{fig:diagram}
\end{figure}

The imaginary part of $\varPi^\text{R}_\sigma(\omega,\bm{p})$ 
is proportional to the difference between the decay and 
creation rates of the mesonic excitations.
For $\bm{p}=\bm{0}$, it is calculated to be
\begin{align}
{\rm Im} \varPi^\text{R}_\sigma( \omega,\bm{0} )
= -\frac 2\pi (
I_\text{PP}(\omega) + I_\text{PC}(\omega) + I_\text{CC}(\omega) ),
\end{align}
with
\begin{widetext}
\begin{align}
I_\text{PP}( \omega )
=& 
- \frac{ p^2 Z_+^2 }{ 2 |\omega_+'| } [ 1-2f_+ ]
\big|_{\omega=-2\omega_+ }
+ \frac{ p^2 Z_-^2 }{ 2 |\omega_-'| } [ 1-2f_- ]
\big|_{\omega=2\omega_- }
- \frac{ 2p^2 Z_+ Z_- }{ |\omega_+'  - \omega_-'| }
[ f_+ - f_- ]
\big|_{\omega = \omega_- - \omega_+ } -(\omega\leftrightarrow-\omega),
\label{eq:imPP}
\\
I_\text{PC}( \omega )
=& 
2\pi\int \frac{d^3k}{(2\pi)^3} Z_+ [ f(\omega_+) - f(\omega+\omega_+) ] 
\rho_-^\text{L} ( \omega+\omega_+ ,\bm{k} ) 
+2\pi\int \frac{d^3k}{(2\pi)^3} Z_- [ f(\omega_- -\omega) - f(\omega_-) ] 
\rho_+^\text{L} (  \omega_- - \omega  ,\bm{k} )\nonumber\\ 
&-(\omega\leftrightarrow-\omega),
\label{eq:imPC}
\\
I_\text{CC}(\omega)
=&
\frac{1}{2}\int \frac{d^3k}{(2\pi)^3} \int dE [ f(E) - f(\omega+E) ]
\rho_+^\text{L}(E,\bm{k}) \rho_-^\text{L} ( \omega+E, \bm{k} ) 
-(\omega\leftrightarrow-\omega),
\label{eq:imCC}
\end{align}
\end{widetext}
with $f(E) = [ \exp(E/T)+1 ]^{-1}$, $f_\pm = f(\omega_\pm)$,
 $\omega'_\pm=d\omega_\pm(p)/dp$,
and each $p$ in Eq.~(\ref{eq:imPP}) denotes the momentum 
of normal quasiquarks or plasminos 
which satisfies $\omega=-2\omega_+$, $\omega=2\omega_-$ and 
$\omega=\omega_- -\omega_+$, respectively.
The first (second) term in Eq.~(\ref{eq:imPP}) includes
the decay process into the normal quasi-quark and quasi-antiquark 
(the plasmino and anti-plasmino), and 
the third term is the Landau damping of the mesons
between the normal quasiparticle and the plasmino.
The decay processes corresponding to 
$I_\text{PC}(\omega)$ and $I_\text{CC}(\omega)$ 
include the continuum $\rho^\text{L}(\omega,\bm{p})$, i.e.
the Landau damping of the quark.
We show the diagrammatical interpretation of decay processes 
included in $I_\text{PC}(\omega)$ and $I_\text{CC}(\omega)$ 
in Fig.~\ref{fig:diagram}(a) and (b), respectively:
Notice that the Landau damping is the scattering processes 
accompanied with a thermally excited particle.

\begin{figure}[tbp]
\begin{center}
\includegraphics[width=.49\textwidth]{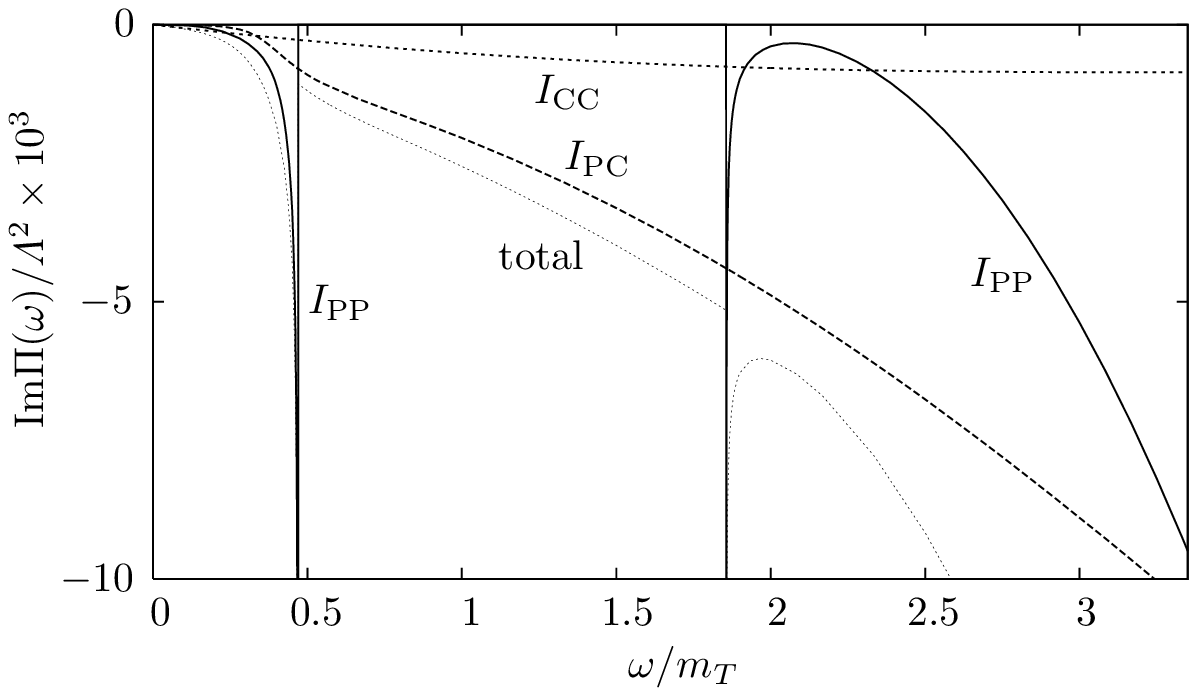}
\caption{
Each component of Im$\varPi^\text{R}(\omega,\bm{p}=\bm{0}) 
= I_{\rm PP}(\omega) + I_{\rm PC}(\omega) + I_{\rm CC}(\omega)$
at $T/\varLambda = m_T/\varLambda = 0.3$.
$I_{\rm PP}(\omega)$, $I_{\rm PC}(\omega)$ and $I_{\rm CC}(\omega)$ 
include the decay processes into
two quasiparticle poles, a quasiparticle pole and continuum, and
two continuums, respectively.
}
\label{fig:imag}
\end{center}
\end{figure}

In Fig.~\ref{fig:imag}, we show each component of Im$\varPi^\text{R}$
for a typical parameter set $ T/\varLambda = m_T/\varLambda=0.3 $.
From the Figure, one sees that $I_\text{PP}(\omega)$ has divergences 
at two energies $\omega_1$ and $\omega_2$, 
with $\omega_1/m_T \simeq 0.4$ and $\omega_2/m_T \simeq 1.8$.
These singularities are due to the divergence of 
the density of states of the energy spectra $2\omega_-(p)$
and $\omega_-(p) - \omega_+(p)$ at finite momenta, 
and interpreted as the van Hove singularities
\cite{BPY90,Karsch:2000gi,Peshier}.
While $I_\text{PP}(\omega)$ vanishes at 
$\omega_1 < \omega < \omega_2$ due to the kinematics,
$I_\text{PC}(\omega)$ and $I_\text{CC}(\omega)$ take finite values 
in the whole range of energy.
In particular, $I_\text{PC}(\omega)$ has a large contribution at
$\omega\lesssim 3m_T$.

\begin{figure}[tbp]
\includegraphics[width=.49\textwidth]{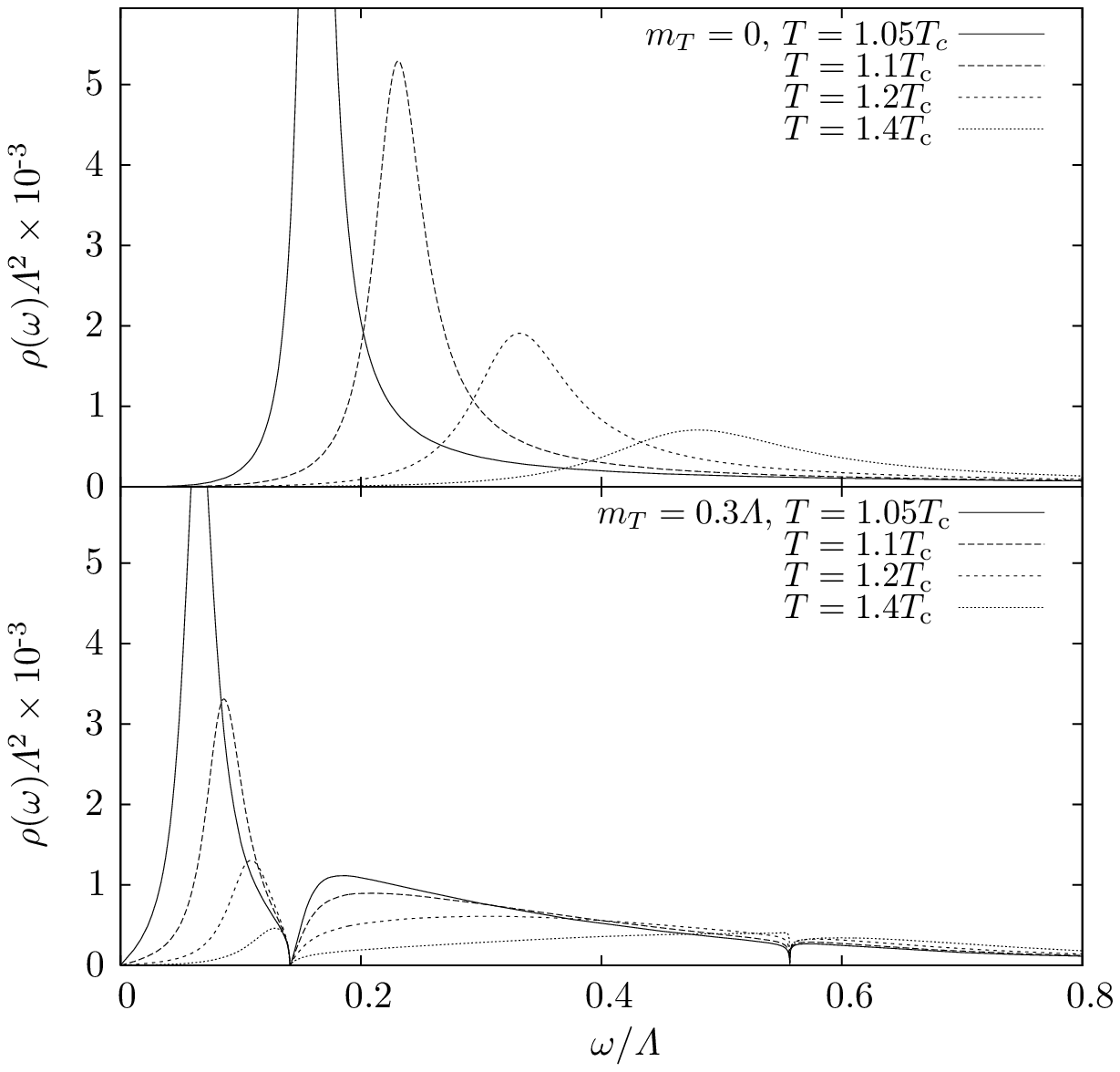}
\caption{
Spectral functions for $m_T=0$ (upper) and $m_T/\varLambda=0.3$ (lower)
with several values of $T$ above $T_c$. 
The coupling $G_\text{S}$ is determined so that $T_\text{c}/\varLambda=0.3$
for each $m_T$.}
\label{fig:spectrum}
\end{figure}

The excitation spectrum of the mesons is given by
the spectral function
\begin{align}
\rho_\sigma( \omega,\bm{p} )
= -2\text{Im} D^\text{R} ( \omega,\bm{p} ).
\end{align}
In Fig.\ref{fig:spectrum}, we show $\rho_\sigma(\omega,\bm{p})$
for $m_T/\varLambda=0$ and $0.3$ at $\bm{p}=\bm{0}$
in the upper and lower panels, respectively; the 
temperature dependence of $m_T$ is not considered as before.
In these Figures, the scalar coupling $G_\text{S}$ is adjusted
so that $T_\text{c}/\varLambda = 0.3$ for each $m_T$.
In the upper panel, we see that there appears a peak 
in $\rho_\sigma(\omega,\bm{p})$, 
and the peak becomes sharper and moves toward the origin
as $T$ approaches $T_\text{c}$ from high temperature.
This is the soft mode of the chiral phase transition \cite{HK85}.

In the lower panel with finite $m_T$, 
$\rho_\sigma(\omega,\bm{p})$ behaves in a more complicated way.
First, 
$\rho_\sigma(\omega,\bm{p})$ becomes zero at two energies
$\omega_1$ and $\omega_2$ corresponding to van Hove singularities
in $I_\text{PP}(\omega)$.
Second, 
although there appears a sharp peak in $\rho_\sigma(\omega,\bm{p})$
as a soft mode of the chiral transition, 
the heights of the peaks in $\rho_\sigma(\omega,\bm{p})$ 
are depressed compared from the upper panel with $m_T=0$.
This result contradicts the na\"{\i}ve expectation that
the mesonic excitation at $\omega<2m_T$ becomes more tightly bound
and thereby have a small width in this range.

\begin{figure}[tbp]
\begin{center}
\includegraphics[width=.49\textwidth]{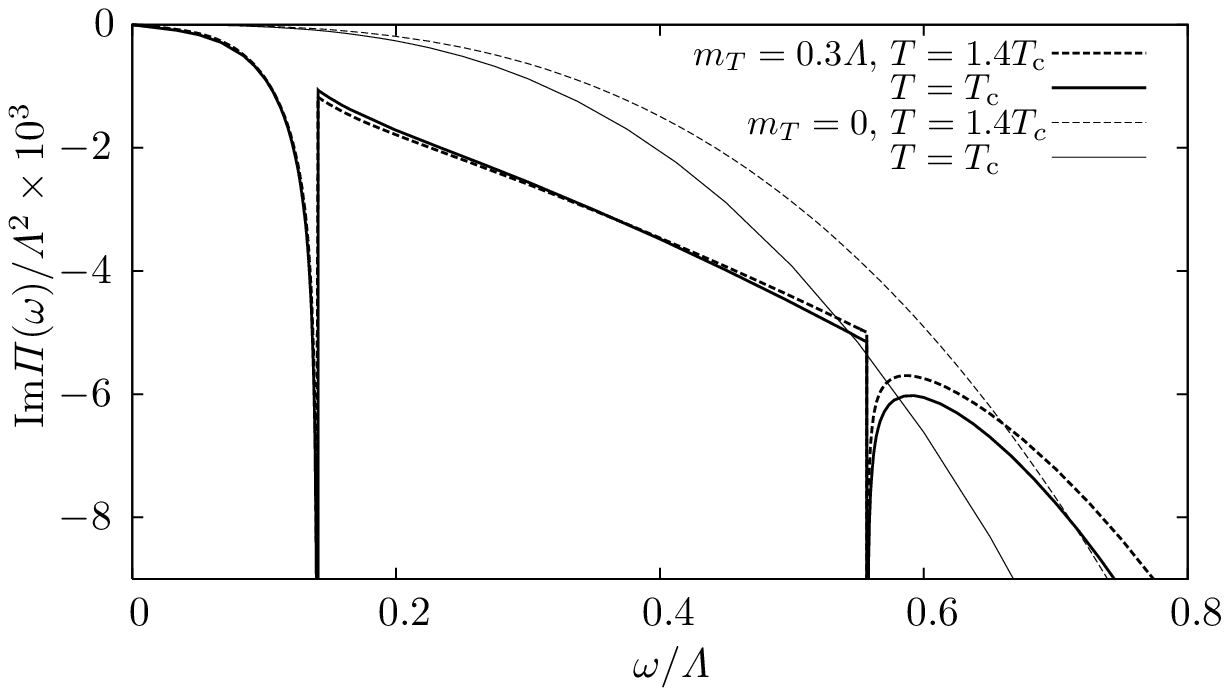}
\caption{
Imaginary part of 
$\varPi^\text{R}(\omega,\bm{p}=\bm{0})$
at $m_T/\varLambda = 0$ and $0.3$ with $T=T_\text{c}$ and $1.4T_\text{c}$.
After including $m_T$, the imaginary part is 
enhanced at $\omega<2m_T$.
}
\label{fig:selfe}
\end{center}
\end{figure}

In order to understand the origin of the large decay width of 
the mesonic excitations with $m_T>0$,
we show Im$\varPi_\sigma(\omega,\bm{p}=\bm{0})$
for $m_T/\varLambda=0$ and $0.3$ in Fig.~\ref{fig:selfe}.
One sees that the imaginary part with $m_T/\varLambda=0.3$ takes 
larger values at $\omega \lesssim 2m_T$ than that
with $m_T=0$, indeed.
As discussed previously,
the decay processes corresponding to $I_\text{PC}(\omega)$ and the
van Hove singularities in $I_\text{PP}(\omega)$ 
has a large contribution in this range.

\section{Summary and Discussions} \label{sec:discussion}

In this work, we explored the effect of the thermal mass of 
the quark on the chiral transition and the mesonic excitations
at finite temperature using a simple model.
It is shown that the order of the chiral transition does not change
by incorporating a thermal mass.
We also found that the thermal mass reduces the chiral condensate.
By studying the spectral function
of the mesonic excitations in the scalar and pseudoscalar channels,
we found that the peak of excitations in the spectral function
is strongly suppressed below $2m_T$.
We showed that the continuum in the quark spectrum, 
and the van Hove singularities, play a significant role 
to enhance the decay rate of the mesonic excitations below $2m_T$.

The decay processes included in $I_{\rm PC}(\omega)$ 
and $I_{\rm CC}(\omega)$
contain the effect of the Landau damping of the quarks, 
i.e. the scattering process of the quasiparticles by 
a thermally excited particle as shown in Fig.~\ref{fig:diagram}.
These processes are specific to nonzero $T$ and
it is natural that such processes are obtained in our calculation
with $S^{\rm HTL}_{m_T}(\omega,\bm{p})$.
Since the contribution of these terms, as well as 
the van Hove singularities, may not depend on 
the channels of the excitation qualitatively,
our calculation strongly suggests that
any mesonic excitations in the light quark sector are
easily destroyed by the thermal effect in the QGP phase.
It is notable that there can nevertheless appear 
sharp peaks in the mesonic spectra 
as the soft modes of the chiral transition near $T_c$ \cite{HK85}.

Although we employed the fixed $m_T$ to show the numerical results,
$m_T \sim gT$ may vary as a function of $T$.
The energies $\omega_1$ and $\omega_2$, i.e. zeros of 
$\rho_\sigma(\omega,\bm{p})$, thus, change with respect to $T$.
Moreover, the width of the collective excitation of the quark 
can depress the van Hove singularities.
Therefore, the depression of the meson spectra is not,
unfortunately, an experimental signature.

While we employed the quark propagator Eq.~(\ref{eq:S})
in this work,
the quark spectrum near $T_\text{c}$ may have the large decay width
and a complicated dispersion \cite{Kitazawa:2005mp,BBS92}.
It is interesting to explore mesonic excitations
with such quark spectra instead of $S^{\rm HTL}_{m_T}$.
This is beyond the scope of this work.

The authors are grateful to Rob.~Pisarski for critical reading of 
the manuscript and valuable comments.
Y.~H. thanks M.~Harada, S.~Yoshimoto and Y.~Nemoto for useful comments.
M.~K. thanks T.~Kunihiro for warm encouragements.
The authors also acknowledge K.~Fukushima, L.~McLerran and A.~Mocsy 
for useful conversations.
The authors are supported by Special Postdoctoral Research Program
of RIKEN.

\end{document}